# Turning Multidimensional Big Data Analytics into Practice: Design and Implementation of *ClustCube* Big-Data Tools in Real-Life Scenarios

Alfredo Cuzzocrea*
iDEA Lab
University of Calabria
Rende, Italy
and Department of Computer Science
University of Paris City
Paris, France
alfredo.cuzzocrea@unical.it

Abderraouf Hafsaoui
iDEA Lab
University of Calabria
Rende, Italy
ahafsaoui.idealab.unical@gmail.com

Ismail Benlaredj
iDEA Lab
University of Calabria
Rende, Italy
ibenlaredj.idealab.unical@gmail.com

*Abstract*—**Multidimensional** Big Data *Analytics* is an emerging area that marries the capabilities of *OLAP* with modern *Big Data Analytics*. Essentially, the idea is engrafting multidimensional models into Big Data analytics processes to gain into expressive power of the overall discovery task. *ClustCube* is a state-of-the-art model that combines OLAP and *Clustering*, thus delving into practical and well-understood advantages in the context of real-life applications and systems. In this paper, we show how *ClustCube* can effectively and efficiently realizing nice tools for supporting Multidimensional Big Data Analytics, and assess these tools in the context of real-life research projects.

## I. INTRODUCTION

*Big Data Analytics* (e.g., [17,18]) has imposed itself as one of the disruptive data technologies of the last decades. Indeed, the number of application scenarios where Big Data Analytics has a relevant impact is paramount. Within this wide and various context, *Multidimensional Machine Learning* (e.g., [1,15,16, 20]) is emerging as one of the key features in the whole Big Data Analytics landscape. The main idea here consists in applying well-consolidated ML methodologies and methods to the core layer of well-understood big data analytics tasks directly, with the advantage of improving expressive power of analytical models and accuracy of retrieved results.

Within this broad context, the OLAP paradigm [11] is a reference pillar, and it represents the theoretical and methodological foundation of the so-called *Multidimensional Big Data Analytics* trend, an emerging trend in the Big Data era [8,9,10]. The main assertion of Multidimensional Big Data Analytics theory relies in the fact that real-life datasets, and, especially, big datasets, are *inherently-multidimensional in nature* [6,19].

In this paper, we show how the state-of-the-art *ClustCube* framework [13], which predicates the marriage between OLAP and *Clustering methodologies*, can be successfully used and exploited for effectively and efficiently supporting Multidimensional Big Data Analytics in real-life big data applications and systems. More into details, we devise and assess via implementations and case studies an *innovative ClustCube methodology in the context of Multidimensional Big Data Analytics*, and clearly prove its feasibility and reliability in real-life tools.

The remaining part of this paper is organized as follows. Section 2 is devoted to a summarized presentation of *ClustCube*. In Section 3, we present in details the *ClustCube* methodology when contextualized in Multidimensional Big Data Analytics settings. Section 4 provides a wide and comprehensive proof-of-concept of the proposed methodology in the *tourism sector*. In Section 5, we introduce a real-life demo of the proposed framework. Finally, in Section 6 we provide conclusions and future work of our research.

## II. CLUSTCUBE IN A NUTSHELL

*ClustCube* [7] represents a *cutting-edge* solution for integrating data from heterogeneous sources which vary both in content and format. This approach facilitates intelligent analysis and predictive processing in the tourism and cultural sectors. It is important to note that data extracted from a distributed database can take significantly different forms than the original stored information. This diversity may be particularly evident in complex objects which could be generated by complex *SQL* instructions involving multiple *JOIN* queries on distributed relational tables.

To fully understand the potential of the extracted data it is crucial to leverage a specialized *Data Mining* component dedicated to extracting these complex objects. This allows for extracting richer and more meaningful information from applications characterized by a high volume of data. The use of advanced Data Mining techniques can lead to considerable benefits, enabling in-depth data analysis and providing valuable insights to improve strategies in the tourism and cultural sectors. In the vast family of available Data Mining techniques, since objects essentially aggregate low-level fields which in turn are extracted attribute values of the distributed database into complex classes, it is natural to think of *Clustering* as the most suitable technique for extracting such derived structures.

Clustering provides an effective method for identifying patterns and hidden structures in the data by allowing for grouping similar objects together based on their common characteristics. This ability to detect correlations and similarities between data makes Clustering a fundamental tool in distributed data analysis.

---

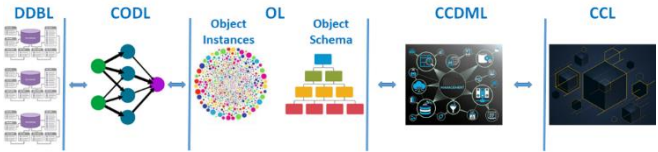

Fig. 1. *ClustCube* Model.

Furthermore, the combined action of Clustering techniques and well-established methodologies developed in the context of OLAP clearly offers powerful tools for extracting clustered objects according to a multidimensional and multi-resolution view of the domain. The integration of these two approaches enables the exploration of data from different perspectives allowing analysts to gain a deeper understanding of the underlying phenomena.

*ClustCube* model combines the power of clustering techniques on complex database objects with the versatility of OLAP in supporting multidimensional analysis and knowledge discovery from grouped complex database objects with mining opportunities and expressive power impossible for traditional methodologies.

In *ClustCube* model, data cubes, complex database objects reside within the data cube cells providing an innovative perspective compared to conventional *SQL-based aggregations* typical of standard *Business Intelligence-oriented* OLAP data cubes. This storage mode allows for more direct and detailed access to grouped objects enabling users to explore and analyze data in deeper and more meaningful ways, according to similar experiences (e.g., [4]). Furthermore, the *ClustCube* approach allows for greater flexibility in identifying complex relationships and patterns within the data by providing analysts with a more sophisticated tool to uncover hidden insights and trends. With the combination of clustering and OLAP *ClustCube* model represents a significant advancement in the field of data analysis offering new perspectives and advanced analytical capabilities to address complex challenges in data management and interpretation.

*ClustCube* defines a reference architecture comprising distinct layers: (i) *Distributed DataBase Layer* (DDBL) housing the target distributed database from which complex objects are extracted; (ii) *Complex Object Definition Layer* (CODL) supporting primitives and functionalities to construct and manage complex objects extracted from the DDBL layer; (iii) *Object Layer* (OL) where complex objects reside along with an appropriate object schema; (iv) *ClustCube Definition and Management Layer* (CCDML) supporting primitives and functionalities for the definition and management of data cube *ClustCube*; (v) *ClustCube* Layer (CCL) storing the final data cubes *ClustCube*. The schema of the *ClustCube* model is depicted in Figure 1, where the different layers of the framework are shown.

This layered architecture ensures a *well-organized structure* of the *ClustCube* model with each layer playing a specific role in the extraction, definition, and management of complex objects and their representation within data cubes *ClustCube*. *Distributed DataBase Layer* serves as the foundational source, *Complex Object Definition Layer* facilitates the creation and handling of complex objects, Object Layer houses the resulting complex objects, *ClustCube Definition and Management Layer* governs the definition and oversight of data cubes *ClustCube* and the *ClustCube* Layer stores the finalized data cubes *ClustCube*. This systematic approach enhances clarity and efficiency in utilizing the *ClustCube* model for advanced data analysis.

Creation and management of complex database objects extracted from the distributed database located at the DDBL level (see Figure 1) play a fundamental role within the *ClustCube* framework. Therefore, in this Section we focus on this aspect, which is based on well-established software methodologies compared to data-intensive applications but tailored to the specific requirements outlined by *ClustCube*. From the DDBL level of the *ClustCube* framework, the CODL level significantly extracts complex database objects used to populate the OL level. This functionality is made possible by a set of *Complex Object Definition Queries* (CODQ). Conceptually CODQ queries reside at the CODL level and are defined by the administrator who modifies the entire *analysis/extraction* process based on their *analysis/extraction* activities. CODQ queries are combined with the CODL level into a singleton global CODQ query indicated by QCODL, which implies a singleton *class/object* schema indicated by OSCODL to which the set of object instances adheres indicated by OICODL. Both the OSCODL schema and the OICODL set are located at the OL level (see Figure 1).

In the *ClustCube* framework, the administrator has the power to modify the global CODQ queries QCODL arbitrarily based on complex SQL expressions defined by the user. This high degree of flexibility allows the administrator to tailor the queries to the specific needs of data analysis, by enabling the exploration of a wide range of scenarios and complexity in information extraction. CODQ queries can be conceived as standard SQL queries; however, their complex nature involves multiple JOIN statements on distributed relational tables located at the DDBL level. This mechanism offers a broad spectrum of possibilities for defining complex objects that do not necessarily correspond to singleton tuple stored in singleton local databases at the DDBL level. This approach opens the door to new and more sophisticated data analysis, by allowing exploration of more intricate relationships and structures within distributed data.

Furthermore, the use of complex queries in the extraction process allows defining mining tasks that would otherwise not be achievable with traditional mining methodologies. This significantly expands the field of analytical possibilities, by enabling the discovery of deeper and more revealing insights in distributed data. Finally, the ability to manipulate CODQ queries in the *ClustCube* framework is one of the fundamental pillars of the methodology, by allowing users to explore and analyze data in *highly customized* and sophisticated ways with significant advantages over conventional mining paradigms.

A *ClustCube* data cube, unlike a traditional data cube, stores complex objects grouped within data cube cells rather than *SQL-based aggregations*. This distinctive feature gives *ClustCube* data cube a unique ability to represent and analyze complex data by allowing for a richer and more detailed multidimensional view. While in traditional OLAP data cubes, the multidimensional boundaries of data cube cells are determined by the input OLAP *aggregation schema*, in data cubes *ClustCube* the multidimensional boundaries of data cube cells are directly the result of the clustering algorithm itself. This means that the relationships and structures present in the data are directly reflected in the arrangement of the cells within the *ClustCube* data cube providing a more faithful representation of the intrinsic complexity of the data.

Furthermore, in traditional OLAP, the *cuboid lattice* associated with an *N-dimensional* data cube is a hierarchical structure composed of cuboids that aggregate original relational data according to arbitrary combinations of dimensions. Each cuboid represents an *n-dimensional view* of the data cube, by allowing for detailed analysis from various perspectives (e.g., [14]). This concept can be further extended in data cubes *ClustCube* where the nature of complex data allows for greater richness of detail and analysis possibilities. In summary, *ClustCube* data cubes represent a significant evolution over traditional data cubes by offering a more accurate and dynamic representation of complex data and opening up new opportunities for advanced analysis and extraction of meaningful insights, with a plethora of interesting applications in emerging big data scenarios.

On the other hand, the *ClustCube* model opens the door to innovative methods for supporting machine learning data analytics, by engrafting advanced methodologies such as *Multidimensional Clustering* (e.g., [5]) and *Multidimensional Regression* (e.g., [3]).

## III. The *ClustCube* Methodology in the Context of Multidimensional Big Data Analytics

In the context of *Multidimensional Big Data Analytics*, the aspect of *multidimensional cuboids* plays a fundamental role. These cuboids, each of which represents an entity in the multidimensional space of data provide a structured and comprehensive view of the data itself. It is important to emphasize that cuboids can vary significantly depending on the dimensions chosen and the level of detail applied to each of them.

The use of cuboids allows for a more in-depth analysis and understanding of the data by enabling the identification of complex relationships and patterns within them. This feature offers enormous potential to further optimize the entire process related to the creation, selection, and management of cuboids. Starting from the *Cuboid Lattice Schema*, it is possible to significantly improve the phases involved in *Data Mart* development. This process includes several crucial stages:

- *Analysis and Reconciliation of Data Sources*: in this phase, it is essential to carefully examine the available data sources by ensuring consistency and integrity before proceeding further;
- *Requirements Analysis*: fully understanding the needs and requirements of the system is fundamental by ensuring all end-user needs are met;
- *Conceptual Design*: key concepts and data relationships are outlined here creating a conceptual structure that will serve as a foundation for subsequent phases;
- *Loading and Validation of the Conceptual Schema*: this phase involves the practical implementation of the conceptual schema, verifying its effectiveness and consistency with the initial requirements;
- *Logical Design*: conceptual concepts translate into a logical model by defining tables, relationships, and

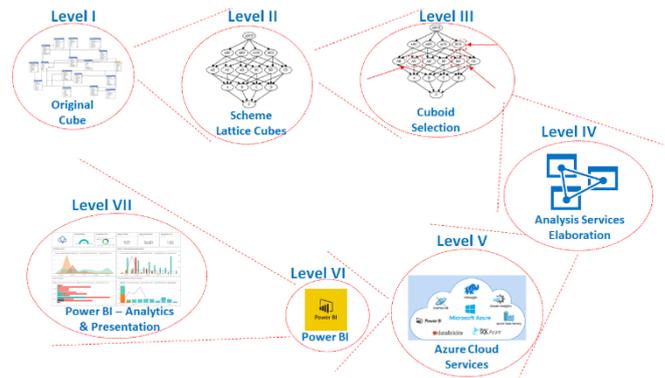

Fig. 2. *ClustCube* Methodology.

  necessary *primary/foreign keys* for system implementation;
- *Physical Design*: technical and infrastructural aspects of system implementation are defined here, optimizing performance and scalability;
- *Feeding Design*: this phase focuses on defining processes for feeding data into the system by ensuring an efficient and reliable flow.

Utilizing the power of Big Data enables these phases to be addressed more efficiently and effectively. The *ClustCube methodology,* applied in this context, is outlined in Figure 2, providing a visual and structured guide for implementing this innovative approach to data management.

*ClustCube* methodology is divided into seven phases or levels as shown in Figure 2. Below, we provide a detailed description of the operations carried out at each level, by using a specific data cube, called *TourismDC*, which is a tourism-sector data cube that will be described deeply in Section 4.1.

Level one represents the starting point. At this level, we can observe the presence of the initial data cube, *TourismDC*. *TourismDC* consists of the *fact table* named *Reservation* to which sixteen dimensions are connected, expressing the analysis dimensions of the model. Among the main dimensions, we observe *Accommodation*, *Point of Interest, Car Rental, Flight, Ferry, Taxi, Tour* and *Tourist*.

Second level implies the configuration of the Cuboid Lattice Schema relating to a data cube. Due to considerations regarding graphical representation, the displayed Cuboid Lattice Schema diagram actually corresponds to a *four-dimensional* data cube rather than the data cube *TourismDC*. This is because the Cuboid Lattice Schema of the data cube *TourismDC* would be too extensive to be displayed within such a schema.

However, the main issue concerns the intrinsic concept of the Cuboid Lattice Schema, which involves representing, in general, the concept that starting from a *fact table* connected to *n dimension tables*, it is possible to create a lattice of cuboids each of which represents a different degree of consolidation along one or more dimensions. In this context, therefore, level two of the schema indicates the representation of cuboids derived from the main data cube *TourismDC*.

At the third level, the cuboid grid is still present, but three cuboids highlighted in red are noticeable indicating they have been selected. This highlighting reflects the selection process carried out by *ClustCube*, which can select cuboids belonging to different levels. This selection enables us to identify cuboids deemed significantly relevant for the subsequent analysis phase of the grid. Thanks to this targeted selection, we care-fully choose cuboids capable of providing optimal results for our objective.

Level four presents the *Analysis Service Elaboration* service. Through the Analysis Service Elaboration operation, it is possible to process Analysis Services elements such as *tabular models*, *data cubes*, *dimensions* and *data mining models*. Multiple elements can be processed simultaneously, either sequentially or concurrently. If there are no elements requiring a specific processing order, concurrent processing can expedite the process. In the case of concurrent processing of elements, the operation can be configured to automatically determine the number of elements to process concurrently or manually specify this number. Sometimes, when processing analysis elements it is necessary to process elements dependent on them as well. *Analysis Services Processing* operation offers an option to process all dependent elements in addition to the selected ones. Processing performed on selected cuboids provides crucial information on both clusters and *Multidimensional Regression*.

Level five is represented by the Cloud. The *Computational Framework of Multi-dimensional Data Analytics System Support* relies on *Microsoft Azure Cloud* infra-structure. Crucial phases of the complete implementation of *Big Data Analytics* techniques are outlined in the following points:

- *Analysis (Big Data Mining)*;
- Prediction (*Big Data Prediction*);
- Visualization (*Big Data Visualization*).

Peculiarity of each phase lies in the adoption of multidimensional modeling methodologies of Big Data aimed at enhancing the effectiveness and expressive power of all phases within the overall process of Big Data Analytics.

Level six and level seven focus on the visualization and processing capabilities, as well as prediction, of the *PowerBI* component. The framework introduces suitable solutions for big data visualization that utilize advanced graphical tools for accessing Big Data. In particular, the types of visualization components that the framework aims to support are as follows:

- *Built-In Big Data Visualization Components*: these visualization components create classic graphical objects (histograms, scatter plots, pies, etc.) directly from the knowledge extracted from Big Data;
- *User-Defined Data Visualization Components*: these visualization components are graphical objects whose appearance can be defined based on specific analysis goals, either natively or by composing the basic Built-In Big Data Visualization Components.

IV. CLUSTCUBE PROOF-OF-CONCEPT:
THE CASE OF THE TOURISM SECTOR

In this Section, we provide a detailed proof-of-concept of the capabilities of *ClustCube* in a real-life setting represented by the tourism sector. We first introduce the data cube *TourismDC*, the central entity of our proof-of-concept in Section 4.1. Then, in Section 4.2 we focus the attention on interesting multidimensional cuboids we derive from the main data cube. Finally, Section 4.3 presents the *ClustCube Mul-tidimensional Big Data Analytics Suite*, a prototypal demo showing the benefits of the *ClustCube* model.

A. The OLAP Data Cube TourismDC

Conceptual diagram in Figure 3 illustrates the *Dimensional Fact Model* (DFM), through which it is feasible to represent data within the OLAP data cube *TourismDC*. It highlights the dimensions of analysis and the relevant facts related to the representation of the modeled reality.

Fundamental aspects of the model are:

- *Providing support for conceptual design;*
- Creating an environment where users can query intuitively and guided.

Graphical representation of the DFM assumes the form of a *Star Schema*. At the center of the schema lies the *Reservation* table, serving as the *fact table* since it contains various measures enabling significant results. Furthermore, it holds considerable importance in the *decision-making* process. The *Reservation* table contains all bookings made by tourists for available services.

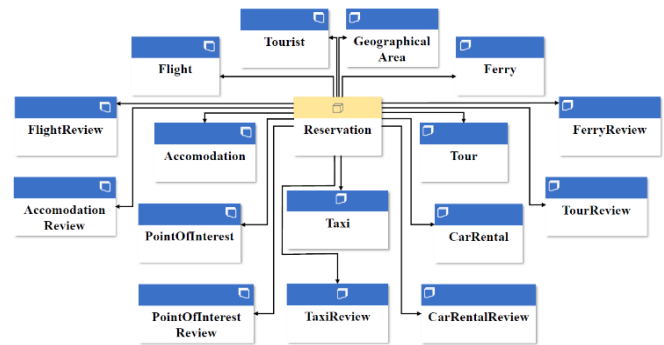

Fig. 3. DFM of the Data Cube *TourismDC*.

Starting from the DFM, the data cube *TourismDC* has been implemented through the integrated environment of *Microsoft Visual Studio 2019* (see Figure 4).

Here, the *fact table* is associated with sixteen *dimensions* and their corresponding relationships that delineate the various analytical dimensions of the model. These dimensions include *Accommodation*, *Point of Interest, Car Rental, Flight, Ferry, Taxi, Tour, Tourist* and *Geographical Area*. Some of these dimensions feature *hierarchies* to allow for a more detailed analysis. In our context, the dimensions represent the services available to tourists. Other dimensions include: *Accommodation Review*, *Car Rental Review, Flight Review, Ferry Review, Tour Review, Taxi Review* and *Point of Interest Review*. The latter dimensions listed, as suggested by their names, represent the evaluations expressed by tourists regarding the quality of the services used.

Figure 4 represents the data cube *TourismDC*, developed using the database *TourismDB*, which contains all the relevant information. Thanks to the integration capabilities provided by software such as *Microsoft Excel, Microsoft SQL Server, SQL Server Management Studio, Microsoft SQL Server Analysis Services* and *Visual Studio 2019,* a path has been traced that starting from the conceptual definition of the schema, through design and implementation, finally led to the creation of the data cube for the Multidimensional Big Data Analytics project.

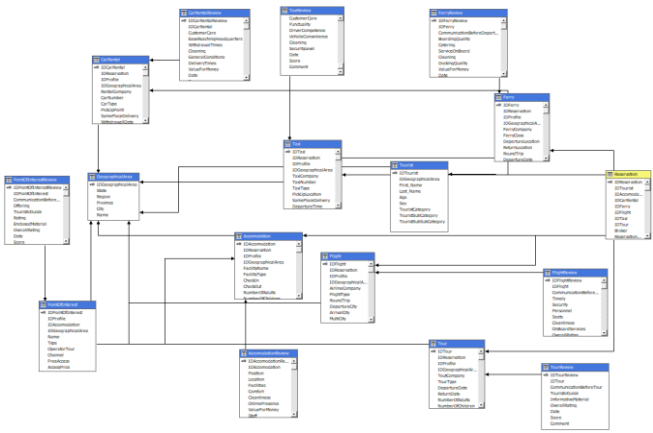

Fig. 4. Data Cube *TourismDC*.

The design activity involves the collection of tourist data, where visitors avail themselves of various services. Exploring the potential of *Multidimensional Analysis* allows for targeted analysis on each variable within the project. This analysis provides a comprehensive and focused view of all aspects related to the variables of interest, such as the implications of a specific choice of tourist service.

Another aspect to underline is the possibility of *real-time data crossing,* meaning having a view of possible scenarios and their variations and implications by simply modifying the variable under observation, thus reflecting changes in the choice directly onto the projection of the reference scenario. Project serves as a valuable investigative tool capable of providing *real-time* answers to all aspects shaping a reality such as tourism and strategic decision-making to better govern a highly complex phenomenon, which plays a fundamental role in the social and economic lives of individuals.

### B. Interesting Multidimensional Cuboids

The *Multidimensional Big Data Analytics* level implements the partitioning of the data cube *TourismDC* to extract important information about specific services. The partitioning involves five cuboids comprised of six dimensions and the *Reservation* table representing the fact table. The five cuboids are respectively named *FlightInformationCube*, *FerryInformationCube*, *CarRentalInformationCube, TourInfor-mationCube* and *TaxiInformationCube*. Each cuboid is oriented towards extracting relevant information for a specific service in the tourism sector.

Challenge and peculiarity of the project lie not only in the innovative elements achieved by *ClustCube* but also in the analysis techniques for processing the information to be obtained, as well as in the concrete and detailed illustration of the *data mining structures* used in the reference software called *PowerBI*. In particular, Multi-dimensional Clustering and Multidimensional Regression are employed.

To provide some example, next we provide the description of two interesting cuboids among all the five we processed.

Cuboid *FerryInformationCube*, shown in Figure 5, consists of the Reservation fact table and six dimensions: *Accommodation*, *Ferry*, *GeographicalArea* and *Tourist*. Dimension *Ferry* is linked to the dimension *FerryReview*, while the dimension *Accommodation* is linked to the dimension *AccommodationReview*. Through the cuboid analysis can be directed concerning the train service offered to tourists. Among the attributes present in the dimension *Ferry*, those indicating railway companies, train types, departure and arrival cities, as well as defined routes, are particularly significant.

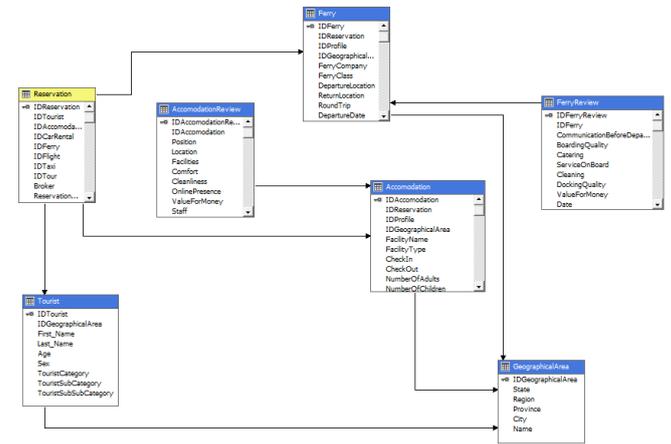

Fig. 5. Cuboid *FerryInformationCube*.

Cuboid *TourInformationCube* in Figure 6 consists of the *Reservation* fact table and six dimensions: *Accommodation*, *CarRental*, *Tourist* and *GeographicalArea*. Dimension *CarRental* is linked to the dimension *CarRentalReview*, while the dimension *Accommodation* is linked to the dimension *AccommodationReview*. Through the cuboid analysis can be directed concerning the depicted services.

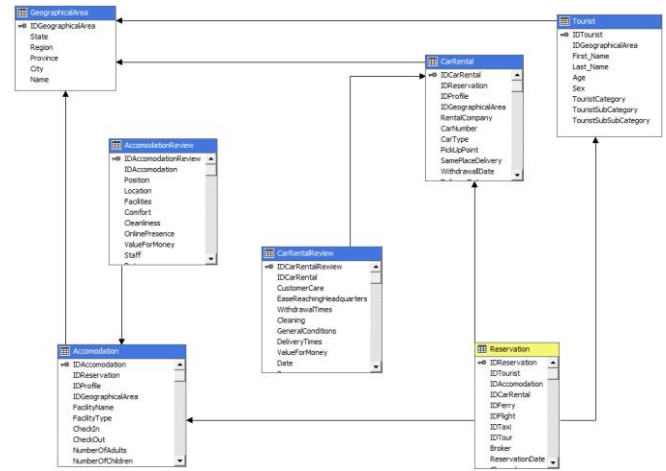

Fig. 6. Cuboid *TourInformationCube*.

## V. THE *CLUSTCUBE* MULTIDIMENSIONAL BIG DATA ANALYTICS SUITE

Work presented in the previous Sections is implemented in a dedicated *demo* called *ClustCube Multidimensional Big Data Analytical Suite.*

Summarizing, this Web suite offers all the functionalities necessary to: (i) access data and metadata of the main data cube and the interesting cuboids; (ii) access the Multidimensional Machine Learning Big Data Analytics tools. All in a nice and user-friendly manner. The suite has been proposed in the context of real-life research projects. Through the homepage, users can access the *Login Page* depicted in Figure 7 by pressing the appropriate button.

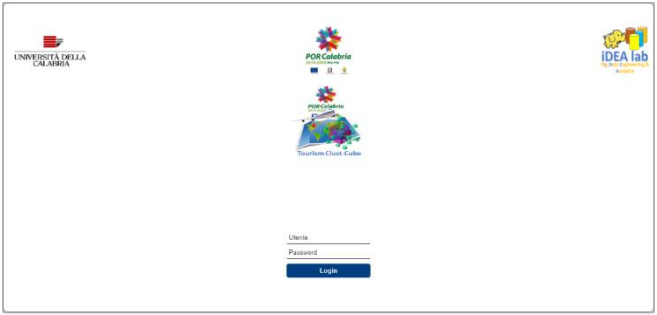

Fig. 7.  Login Page.

After login, the page that appears to the user contains a tree structure starting from the first level called *TourismDB*, which includes a second level called *Tables* that contains the tables within the database *TourismDB*, and *TourismDC*, which when expanded presents the measures *Measures*, the dimensions *Dimensions* of the data cube and the processed cuboids at the second level.

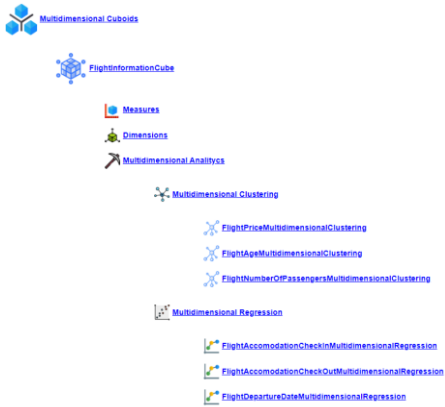

Fig. 8.  Section *Multidimensional Cuboid Exploration*.

As said, the most important aspect of the suite is represented by multidimensional cuboids and analytics over them. In this respect, Figure 8 shows the *Multidimensional Cuboid Exploration* section, which explores the cuboids including measures, dimensions, and analysis of the processed cuboids. These cuboids, which we selected as interesting cuboids in our proof-of-concepts, are: *FlightInformationCube*; *TaxiInformationCube*; *CarRentalInformationCube*; *TourInformationCube*; *FerryInformationCube*. As regards analytics, in Figure 9, the section *Multidimensional Clustering Analysis* is illustrated, while, in Figure 10, the section *Multidimensional Regression Analysis* is shown.

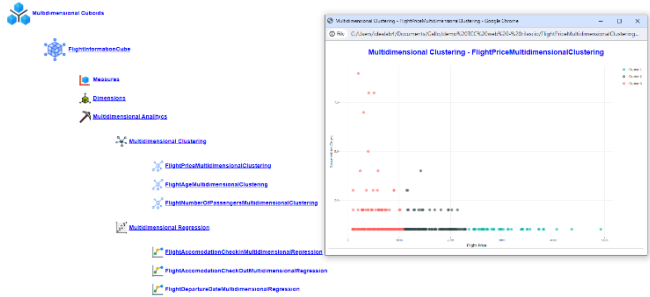

Fig. 9.  Section *Multidimensional Clustering Analysis*.

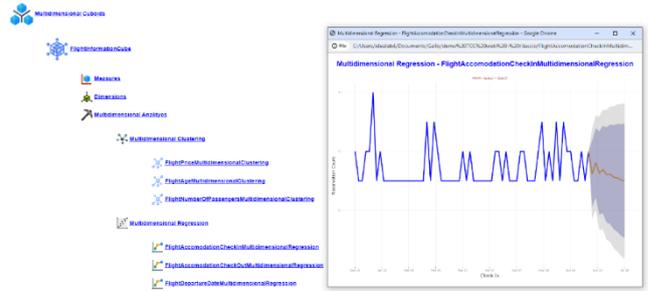

Fig. 10. Section *Multidimensional Regression Analysis*.

## VI. CONCLUSIONS AND FUTURE WORK

In this paper, we have provided our methodology for applying the state-of-the-art *ClustCube* framework to the real-life problem of supporting Multidimensional Ma-chine Learning over Cloud-enabled big data infrastructures, and we have shown its proof-of-concept in the tourism sector. Future work is mainly oriented to engraft into our framework innovative aspects of big data processing (e.g., [2,12,13]).

## ACKNOWLEDGMENT

This work was partially funded by the Next Generation EU - Italian NRRP, Mission 4, Component 2, Investment 1.5 (Directorial Decree n. 2021/3277) - project Tech4You n. ECS0000009.

The authors are grateful to Carmine Gallo and Marco Antonio Mastratisi who participated in an early version of this research.